\documentclass[a4paper,fleqn,usenatbib]{mnras}

\usepackage{newtxtext,newtxmath}

\usepackage[T1]{fontenc}
\usepackage{ae,aecompl}

\usepackage{graphicx}	
\usepackage{amsmath}	
\usepackage{amssymb}	
\usepackage{color}

\title[Co-existence of fast and $\Omega$-slow wind solutions]{Co-existence and switching between fast and $\Omega$-slow wind solutions in rapidly rotating massive stars}

\author[I. Araya et al.]{
I. Araya,$^{1,2}$\thanks{E-mail: ignacio.araya@umayor.cl}
M. Cur\'e,$^{1}$
A. ud-Doula,$^{3,4}$
A. Santill\'an,$^{5}$
L. Cidale,$^{6,7,1}$\thanks{Member of the Carrera del Investigador Cient\'{\i}fico, CONICET, Argentina} 
\\
$^{1}$Instituto de F\'{\i}sica y Astronom\'{\i}a, Facultad de Ciencias, Universidad de Valpara\'{\i}so, Av. Gran Breta\~na 1111, Casilla 5030, Valpara\'{\i}so, Chile\\
$^{2}$N\'ucleo de Matem\'aticas, F\'isica y Estad\'istica, Facultad de Ciencias, Universidad Mayor, Chile\\
$^{3}$Penn State Worthington Scranton, Dunmore, PA 18512, USA\\
$^{4}$Dept A.G.O., Universite de Liege, Belgium \\
$^{5}$Direcci\'on General de C\'omputo y Tecnolog\'ias de la Informaci\'on y Comunicaci\'on, Universidad Nacional Aut\'onoma de M\'exico, \\Ciudad Universitaria, C.P. 04510, Mexico City, Mexico\\
$^{6}$Departamento de Espectroscop\'{\i}a, Facultad de Ciencias Astron\'omicas y Geof\'{\i}sicas,
Universidad Nacional de La Plata (UNLP),\\ Paseo del Bosque S/N, 1900 La Plata, Argentina\\
$^{7}$Instituto de Astrof\'{\i}sica La Plata, CCT La Plata, CONICET-UNLP, Paseo del Bosque S/N, 1900 La Plata, Argentina\\
}

\date{Accepted XXX. Received YYY; in original form ZZZ}

\pubyear{2017}

\begin{document}
\label{firstpage}
\pagerange{\pageref{firstpage}--\pageref{lastpage}}
\maketitle

\begin{abstract}
Most radiatively-driven massive star winds can be modelled with m-CAK theory  resulting in so called fast solution. However, those most rapidly rotating among them, especially when the stellar rotational speed is higher than $\sim 75\%$ of the critical rotational speed, can adopt a different solution called $\Omega$-slow solution characterized by a dense and slow wind. Here, in this work we study the transition region of the solutions where the fast solution changes to the $\Omega$-slow. Using both time-steady and time-dependent numerical codes, we study this transition region for different equatorial models of B-type stars. In all the cases, at certain range of rotational speeds, we found a region where the fast and $\Omega$-slow solution can co-exist. We find that the type of solution obtained in this co-existence region depends heavily on the initial conditions of our models. We also test the stability of the solutions within the co-existence region by performing base density perturbations in the wind. We find that under certain conditions, the fast solution can switch to a $\Omega$-slow solution, or vice versa. Such solution switching may be a possible contributor of material injected into the circumstellar environment of Be stars, without requiring rotational speeds near critical values.

\end{abstract}

\begin{keywords}
hydrodynamics -- stars: winds, outflows -- stars: rotation -- stars: early-type
\end{keywords}

\section{Introduction}

Massive stars are characterized by strong outflowing stellar winds, driven by the transfer of momentum from the radiation field to the plasma via the scattering processes in spectral lines. These winds are called radiation-driven winds and are described by the standard radiation-driven wind theory (or CAK theory), based on the pioneering work of \citet{castor1975}. Modifications to the CAK model (m-CAK) were performed by \citet{friend1986} and \citet{ppk1986} who implemented the finite disk correction factor, which accounts for the cone angle subtended by the star instead of a point star approximation. This m-CAK model has been successful in describing both wind terminal velocities ($v_{\infty}$) and mass-loss rates ($\dot{M}$) from hot stars \citep{puls2008}.

The effects of the stellar rotation on the wind dynamics was also investigated by \citet{friend1986} and \citet{ppk1986}. Both studies found that in general stellar rotation increases mass-loss rate and decreases terminal speed of the wind for slow or moderate rotational speeds. Particularly, \citet{friend1986} were unable to find a wind solution for rotation speeds higher than about $75\%$ of the critical rotational speed. This was a strong indication that something unusual happens to the wind solution for rapidly rotating massive stars.

Among  massive stars there are many rapid rotators, even with rotational speeds near its critical rate. The latter condition has been generally assumed to explain the formation of a circumstellar disk, e.g., Be stars \citep{townsend2004,rivinius2013}.
A 2D model to link radiation-driven wind theory to the formation of a dense thin equatorial disk in Be stars was developed by \citet{bjorkman1993}, called Wind Compressed Disk model (WCD). Then \citet{owocki1994} implemented a time-dependent WCD model with purely radial forces and confirmed the basic WCD effect. However, \citet{cranmer1995} included in this model non-radial line forces  and showed that these forces inhibit the formation of an equatorial disk. This result was also confirmed by \citet{petrenz2000}.

\citet{cure2004} revisited the effects of stellar rotation in the theory of radiation-driven winds for 1D stationary models and found a new type of wind solution when the rotational speed ($v_{\mathrm{rot}}$) is above  $\sim 75\%$ of the critical rotational speed ($v_{\mathrm{crit}}$) and analysed its application to explain the equatorial disk of Be stars. This new solution, called the $\Omega$-slow solution, being $\Omega=v_{\mathrm{rot}}/v_{\mathrm{crit}}$, is slower and denser than the standard solution (hereafter fast solution). This model combined with the bi-stability effect \citep{lamers1991} predicts equator to pole density contrasts up to  $10^{4}$ suggesting that the $\Omega$-slow solution could plays a key role in the formation of dense equatorial disks in B[e] stars \citep{cure2005}.0 With the purpose of a better understanding of this slow wind solutions and its stability at high rotational speeds, \citet{madura2007} calculated time-dependent simulations and concluded that it was unlikely that the $\Omega$-slow solution could reproduce the inferred densities of B[e] disks. In addition, they reported the presence of abrupt kink transitions in the wind velocity profiles at rotation rates around $75\%$--$85\%$ of the critical rotational speed. 

Despite the importance of 2D simulations, 1D calculations represent best case scenario. In this way, this paper aims to study the region where the fast solution switches to the $\Omega$-slow solution (or vice versa) for a 1D equatorial model at high rotational speed. Our calculations are performed using stationary and time-dependent hydrodynamic codes. We also study the dependence of a co-existence region with the stellar and line-force parameters (m-CAK formalism) and the stability of the solutions obtained within this region, using base density wind perturbations. 

The paper is organized as follows: Section \S 2 introduces our hydrodynamic wind model, starting with our assumptions. Then, we outline the time-dependent and steady state equations, used to describe the rotating m-CAK theory. In Section \S 3, we solve the stationary equations in the star's equatorial plane for a wide range of rotational speeds where fast and $\Omega$-slow solutions are achieved. In addition, a co-existence region is found where both types of solutions are present. The influence of the line-force parameters on this regions is also studied. Section \S 4  presents the solutions obtained from time-dependent simulations and a comparison between stationary and time-dependant  results.  In Section  \S 5, we perform  wind base density perturbations, with the purpose to test: i) the stability of both solutions inside the co-existence region and ii) the minimum conditions to induce a regime-switching between wind solutions. Section \S 6 discusses the application of our results. 
Finally, we present our conclusions in Section \S 7.

\section{Hydrodynamic wind model}
\label{approx}
We adopt the time-dependent 1D rotating radiation driven m-CAK model for the equator of a massive star. This model uses the following approximations: {\it{i}}) An isothermal wind with temperature equal to the stellar effective temperature and {\it{ii}}) Angular momentum conservation. This model neglects the effects of: viscosity, heat conduction and magnetic fields. In addition, the distortion of the star's shape caused by its high rotational speed,  gravity darkening, non-radial velocities  and multi-scattering are not considered.

\subsection{Time-dependent equations}

At the equatorial plane, the relevant 1D time-dependent hydrodynamic equations, namely continuity and radial momentum, are:

\begin{equation}
\label{continuity}
\frac{\partial \rho}{\partial t} + \frac{1}{r^{2}}\frac{\partial}{\partial r}(\rho\, r^{2}\,v)=0
\end{equation} 

\noindent and

\begin{equation}	
\label{momentum}
 \frac{\partial v}{\partial t} + v\frac{\partial v}{\partial r}=-\frac{1}{\rho}\frac{\partial p}{\partial r}- \frac{G\,M (1-\Gamma_{\mathrm{E}})}{r^{2}}+\frac{v^{2}_{\phi}}{r} + g^{\mathrm{line}},
\end{equation}

\noindent where $t$ is the time, $r$ is the radial coordinate, $v$ is the fluid radial velocity, $\rho$ is the mass density and $g^{\mathrm{line}}$ is the radiative line acceleration described in more detail below. The ideal gas pressure ($p=a^2 \, \rho$) is given in terms of the isothermal sound speed, $a$. The azimuthal speed $v_{\phi}=v_{\mathrm{rot}}\,R_{*}/r$ is calculated assuming conservation of angular momentum, where $v_{\mathrm{rot}}$ is the star's surface rotational speed at equator and $R_{*}$ is the stellar radius. The classical Eddington factor, expressed as
\begin{equation}
\label{gammaE}
\Gamma_{\mathrm{E}} = \frac{\sigma_{\mathrm{E}}\, L}{4\, \pi \, c \, G \, M},
 \end{equation}

\noindent  is the ratio between Thomson electron scattering force and the gravitational force. The  electron scattering opacity per unit mass  $\sigma_{\mathrm{E}}$ is given by,

\begin{equation}
\sigma_{\mathrm{E}} = \sigma_{\mathrm{Th}} \frac{n_{\mathrm{E}}}{\rho},
\end{equation}

\noindent with $\sigma_{\mathrm{Th}}$ the Thomson cross-section of electrons and $n_{\mathrm{E}}$ the electron number density. In addition, 
\begin{equation}
\label{eq5}
\frac{n_{\mathrm{E}}}{\rho} = \, D\, = \frac{1}{m_{\mathrm{p}}} \left(  \frac{1+Z_{\mathrm{He}}\, Y_{\mathrm{He}}}{1+4\, Y_{\mathrm{He}}} \right),
\end{equation}

\noindent where $m_{\mathrm{p}}$ is the mass of the proton, $Y_{\mathrm{He}} = n_{\mathrm{He}}/n_{\mathrm{H}}$ is the helium abundance relative to hydrogen, and $Z_{\mathrm{He}}$ is the number of free electrons provided by He. 

Following the descriptions of  \citet{abbott1982}, \citet{friend1986}, and \citet{ppk1986}, the m-CAK standard parametrization for the radiative line acceleration term is:

\begin{equation}
\label{gline}
 g^{\mathrm{line}} = \frac{\Gamma_{\mathrm{E}}\,G\,M\,k}{r^{2}} \left( \frac{1}{\sigma_{\mathrm{E}}\,v_{\mathrm{th}}} \right)^{\alpha}  \left( \frac{1}{\rho} \frac{\partial v}{\partial r} \right)^{\alpha} \left( \frac{n_{\mathrm{E11}}}{W(r)} \right)^{\delta} f_{\mathrm{FD}}\left( r,v,\frac{\partial  v}{\partial  r}\right), 
\end{equation}

\noindent where  $v_{\mathrm{th}}$ is the mean thermal velocity of the protons, $W(r)$ the dilution factor and $n_{\mathrm{E11}}$ is the electron number density in units of  $10^{-11}$ cm$^{-3}$.

The m-CAK  line-force parameters \citep{abbott1982,puls2000} are given by $\alpha$ (the ratio between the line-force from optically thick lines and the total line-force), $k$ (which is related to the number of lines effectively contributing to the driving of the wind), and $\delta$ (which accounts for changes in the ionization throughout the wind). 

We include the effects of finite disk correction factor, $f_{\rm{FD}}$, due to the finite size of the star  \citep[see eq. 50 from ][]{castor1975}.





A different approach for the radiative line acceleration was considered by \citet{gayley1995}, who introduced an alternative formalism where the line-force parameter $k$ is replaced by the dimensionless 
line-strength parameter $\bar{Q}$ and the cutoff parameter $Q_{\mathrm{o}}$.  These parameters are related by the following identity:

\begin{equation}
\label{ktoq}
k= \frac{1}{1-\alpha} \left( \frac{v_{\mathrm{th}}}{c} \right)^{\alpha} \, \bar{Q}\, Q_{\mathrm{o}}^{-\alpha}.
\end{equation}

\noindent  Based on the \citet{gayley1995}'s formalism, the $ g^{\mathrm{line}}$ term can be expressed as:
\begin{equation}
\label{gline-gayley}
 g^{\mathrm{line}} = \frac{\Gamma_{\mathrm{E}}\,G\,M\,\bar{Q}}{\left( 1 - \alpha\right) r^{2}} \left( \frac{1}{Q_{\mathrm{o}}\,\sigma_{\mathrm{E}}\,c} \right)^{\alpha}  \left( \frac{1}{\rho} \frac{\partial v}{\partial r} \right)^{\alpha} \left( \frac{n_{\mathrm{E11}}}{W(r)} \right)^{\delta} f_{\mathrm{FD}}\left( r,v,\frac{\partial  v}{\partial  r}\right). 
\end{equation}

Here we follow the original CAK formalism, but conversion to Gayley method can be readily made. \citet{Owocki1988} uses a slightly different formalism by introducing maximum opacity $\kappa_{\mathrm{max}}$ that limits the effects of strong driving lines in their line-deshadowing instability models.

\subsection{Steady state equations}
For our stationary 1D spherical flow, we combine the continuity equation (\ref{continuity})  and momentum equation ( \ref{momentum}) assuming $\partial/\partial t \, \rightarrow 0 $. This leads to our basic equation of motion:

\begin{equation}
\label{motion1}
\left( 1- \frac{a^{2}}{v^{2}} \right) v \frac{dv}{dr} =\frac{2\, a^{2}}{r} - \frac{G\,M (1-\Gamma_{\mathrm{E}})}{r^{2}}+\frac{v^{2}_{\phi}}{r} + g^{\mathrm{line}}.
\end{equation}

To facilitate a solution,  the following changes of variables, $u=-R_{*}/r$, $w=v/a$ and $w'= dw/du$ leads to \citep{castor1975,cure2004}:

\begin{eqnarray}
\nonumber
\label{motion-eq}
F(u,w,w') & \equiv & \left( 1- \frac{1}{w^{2}} \right) w \frac{dw}{du} + A + \frac{2}{u} + a^{2}_{\mathrm{rot}}\, u \\
& &  - \, C' \, f_{\mathrm{FD}} \, g(u)\, (w)^{-\delta} \left( w \frac{dw}{du} \right)^{\alpha} =0, 
\end{eqnarray}

\noindent  where $a_{\mathrm{rot}}=v_{\mathrm{rot}}/a$, $A= \frac{G\, M \left[ 1- \Gamma_{\mathrm{E}}\right]}{a^{2}R_{*}}$,
$g(u)= \left( \frac{u^{2}}{1-\sqrt{1-u^{2}}}  \right)^{\delta}$

\noindent and

\begin{equation}
\label{eigenvalue}
C' = C \left( \frac{\dot{M}\, D}{2\, \pi} \frac{10^{-11}}{a\,R_{*}^{2}}  \right)^{\delta} \left( a^{2}R_{*} \right)^{\alpha-1}.
\end{equation}

\noindent With $C'$ being the eigenvalue of the problem where $C= \Gamma_{\mathrm{E}}\, G \, M \, k \left( \frac{4\pi}{\sigma_{\mathrm{E}} \, v_{\mathrm{th}} \, \dot{M}} \right)^{\alpha}$
and $D$ is given by Equation (\ref{eq5}). \\

Based in this formalism, \citet{cure2004} demonstrated the existence of another physical solution for fast rotating stars.

\section{1D steady state wind solutions for fast rotating stars}
\label{solu}
Solving Eq. \ref{motion-eq} for high rotational speeds, \citet{cure2004} found that the standard (or fast) solution ceases to exist and a new kind of solution is found from $\Omega \gtrsim 0.75$, the so-called $\Omega$-slow solution.
Now we want to explore if there is an interval in the $\Omega$-parameter space where the wind solution switches from fast to $\Omega$-slow solution or vice versa. If this interval exists, it might be possible that both solutions can co-exist.

To obtain wind numerical solutions, we use our stationary hydrodynamic code \textsc{Hydwind} described in \citet{cure2004}. \textsc{Hydwind} solves Eq. \ref{motion-eq} by relaxation method, thus it requires an initial trial solution for the velocity profile and a guess value for the eigenvalue to begin the integration. Usually a $\beta$-law is used as initial trial solution. For fast solutions we use typically $\beta \approx 0.8$ and  $v_{\infty} \approx 1000$ km s$^{-1}$ and for slow solutions $\beta \approx 3.5$ and $v_{\infty} \approx 400$ km s$^{-1}$. The type of solution obtained (fast or $\Omega$-slow) depends solely on the initial trial solution for the velocity profile and not on the base wind density, because the latter is fixed constant independently of the rotational speed.

We begin our analysis calculating wind solutions for $0 \leq\Omega \leq 0.95$ using the stellar parameters for a typical main-sequence B2.5 star following the work of \citet{madura2007}: $T_{\mathrm{eff}}=20$ kK, $R_{*}=4$ $R_{\odot}$ and $\log\mathrm{g}=4.11$. The base density was fixed at $\rho_{0}=8.709 \times 10^{-13}$ g cm$^{-3}$ and the line-force was parametrized using the \citet{gayley1995}'s formalism:  $\alpha=0.5$, $\delta=0.0$ and $\bar{Q}=1533$\footnote{\citet{madura2007} assume the \citet{gayley1995}'s formalism with $\bar{Q} \approx Q_{\mathrm{o}}$, contrary to what  \citet{puls2000} states for $T_{\mathrm{eff}} < 35\,000$ K. Independently of this, we use the same assumption with the purpose to compare results.} (or $k=0.6098$, according to Equation \ref{ktoq}). 

\begin{figure}
	\includegraphics[width=\columnwidth]{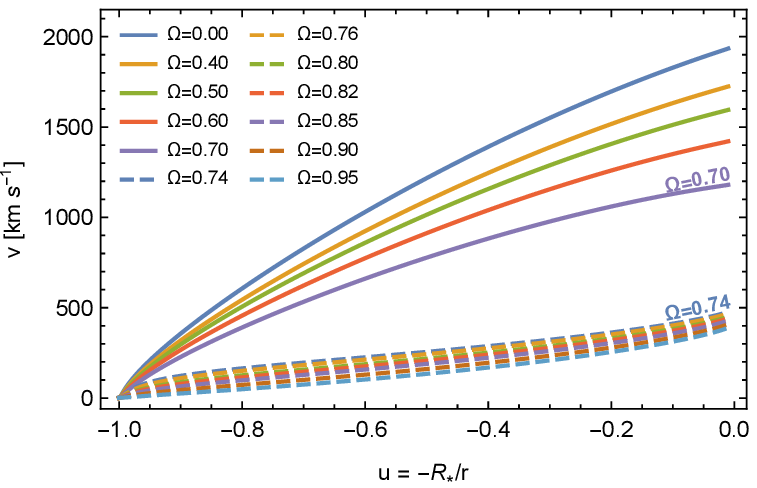}
	\vskip 0.3cm	
	\includegraphics[width=\columnwidth]{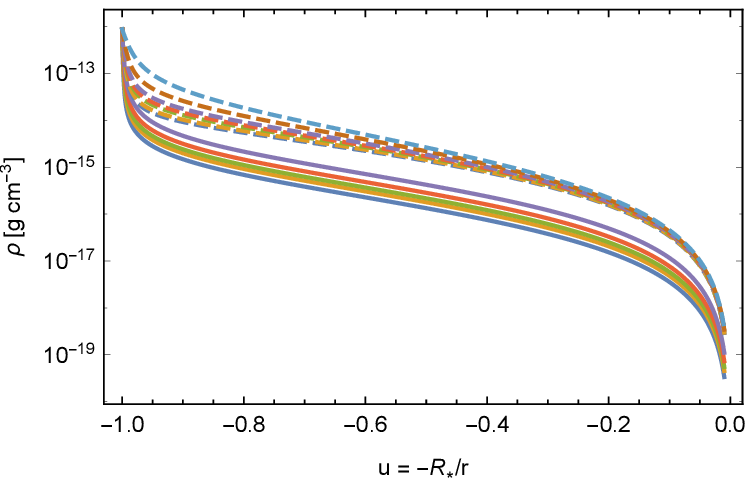}	
    \caption{Steady state wind solutions at different rotational speeds $\Omega$ for a main-sequence B2.5 stellar model. The solutions were obtained with the stationary code \textsc{Hydwind}. Top and bottom panels correspond to the velocity and density profiles as a function of the inverse radial coordinate $u$, respectively. The solid and dashed lines correspond to fast and $\Omega$-slow solutions, respectively.}
    \label{mof-hyd-profiles}
\end{figure}

Figure \ref{mof-hyd-profiles} shows some of our calculated stationary velocity and density profiles. In this figure, we clearly observe that the fast solution regime is characterized by high velocities and low densities, while the $\Omega$-slow solution regime is characterized by lower velocities and higher densities. The switch of regime can be observed at $\Omega \sim 0.74$. 

\begin{figure}
	\includegraphics[width=\columnwidth]{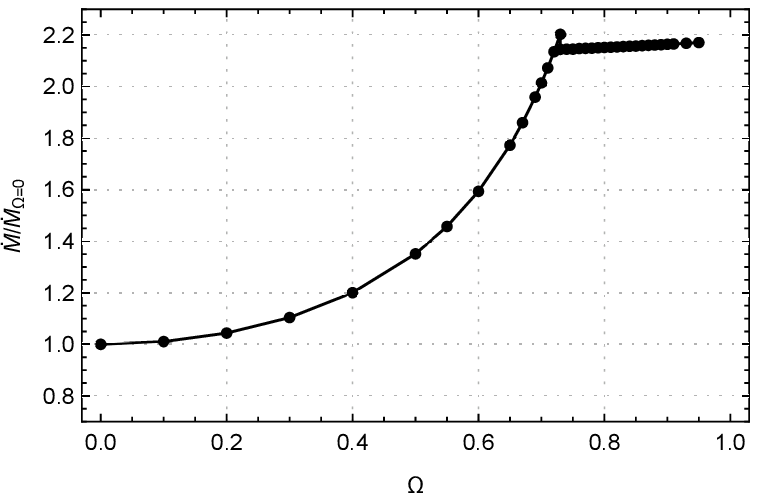}
	\vskip 0.3cm	
	\includegraphics[width=\columnwidth]{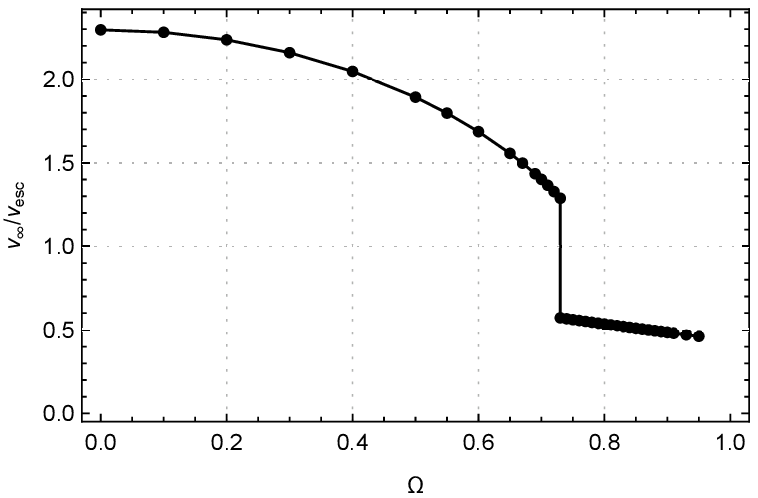}
	\vskip 0.3cm	
	\includegraphics[width=\columnwidth]{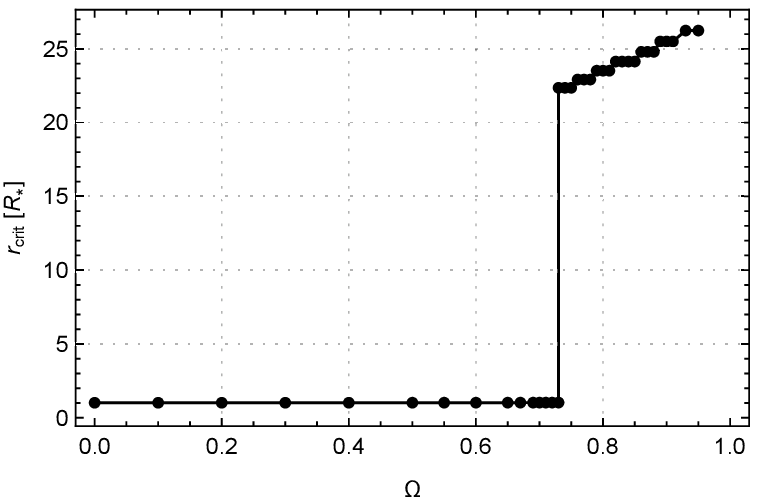}	
    \caption{{ Mass-loss rate (in units of the mass-loss for a non-rotating case, top panel), terminal velocity (in terms of the escape velocity, middle panel) and location of the critical point (bottom panel) as function of the rotational speed. These values were obtained using a hydrodynamical stationary solution with the code \textsc{Hydwind}} using $\dot{M}_{\Omega=0} =1.004\times10^{-9}\, M_{\sun}\mathrm{yr}^{-1}$.
    }
    \label{mof-hyd}
\end{figure}

In Figure \ref{mof-hyd} we show the mass-loss rate (in units of the mass-loss of a non-rotating case), ratio between terminal and escape velocity, and location of the critical point as a function of the rotational speed \citep[equation of motion's topology and integration procedure are explained in detail in][]{cure2004}. The differences between fast and $\Omega$-slow solutions are quite clear.  For $\Omega \sim 0.74$ there is a jump in the terminal velocity and position of the critical point between these solutions, while the mass-loss rate presents a change in its slope.

The velocity profiles shown in Figure \ref{mof-hyd-profiles} as well as the values of the mass-loss rates and $v_{\infty}/v_{\rm{esc}}$ shown in Figure \ref{mof-hyd} are similar to the analytical results given by \citet[][see their figures 3 and 4]{madura2007}. 
In that work, the authors described these solutions as a flow with a steep and shallow acceleration for $\Omega \leq 0.74$ and $\Omega \geq 0.75$, respectively. In Section \ref{time_dep} we will discuss our time-dependent calculations and compare them with the results performed by \citet{madura2007}.

\subsection{Co-existence of fast and slow regimes}
\label{coexistence}
In this section, we focus on the interval in the $\Omega$-space parameter domain where the fast solution ceases to exist and the $\Omega$-slow solution emerges. 

We apply our analysis to three different  models taken from the literature: a main-sequence B2.5 star  (same as previous section), a subgiant B0 star and a supergiant B3 star. The stellar and line-force parameters for these objects are given in Table \ref{param}. 
The B0 IV model parameters are from \citet{sigut2007} and the line-force parameters are derived from \cite{ppk1986}. The stellar and line-force parameters for the B3 I model are taken from the equatorial bi-stable wind model from \citet{pelupessy2000}. 

\begin{table}
\center
 \caption{Stellar and line-force parameters for our  models.}
 \label{param}
 \tabcolsep 2.7 pt
 \begin{tabular}{lccccccrc}
  \hline
  Model & T$_{\mathrm{eff}}$ & $\log g$ & $R_{*}$ & $\alpha$ & $\delta$ & $k$& $\Bar{Q}$ & $\rho_{0}$ \\
       & [kK]                  &  [dex]       & [R$_{\sun}$] & & & & & [g\,cm$^{-3}$]\\
  \hline
  B0 IV   & 25.0 & 3.50 & 10.00 & 0.565 & 0.02 & 0.32& 2792.0 &$5.0 \times 10^{-11}$\\  
  B2.5 V  & 20.0 & 4.11 & 4.00  & 0.500 & 0.00 & 0.61& 1533.0 &$8.7 \times 10^{-13}$ \\
  B3 I    & 17.5 & 2.70 & 47.00 & 0.450 & 0.00 & 0.57& 361.6 &$1.0 \times 10^{-11}$\\
  \hline
 \end{tabular}
\end{table}

\begin{figure}
	\includegraphics[width=\columnwidth]{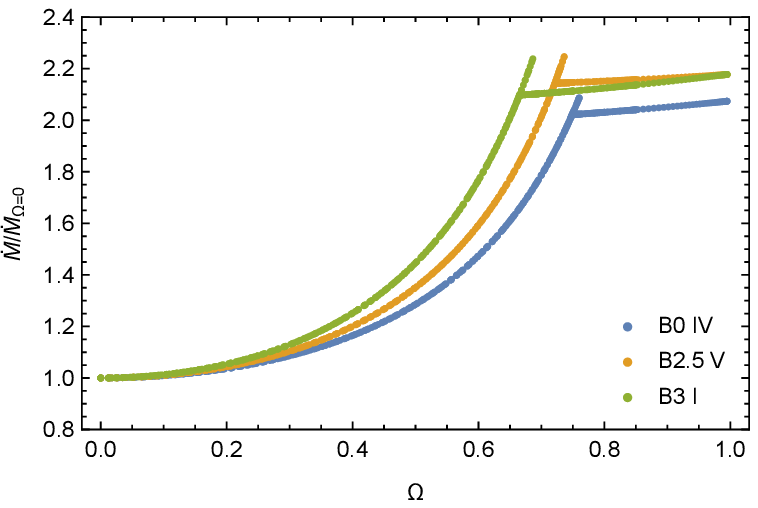}
	\vskip 0.3cm	
	\includegraphics[width=\columnwidth]{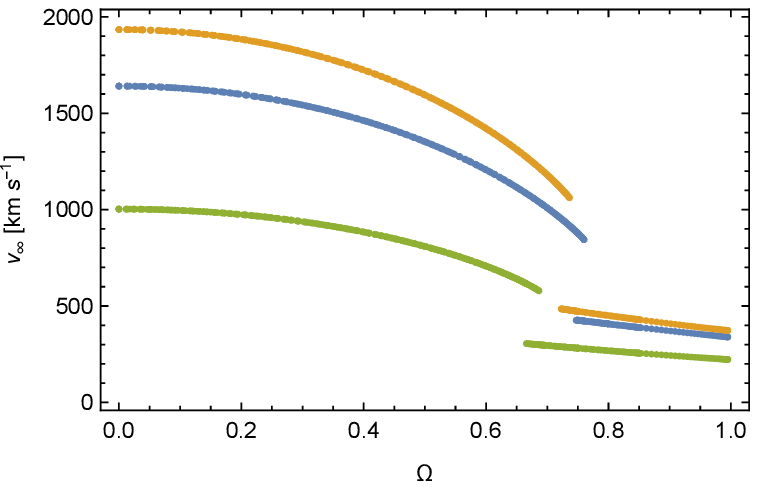}
        \caption{Mass-loss rates in units of a non-rotating case (top panel) and terminal velocities (bottom  panel) as function of $\Omega$ for the solutions obtained from three stellar models. The solutions are calculated with \textsc{Hydwind}. An overlap of fast and $\Omega$-slow solutions is observed in the range $0.65\lesssim \Omega \lesssim 0.75$.}
    \label{stars-coex}
\end{figure}

\begin{figure}
	\includegraphics[width=\columnwidth]{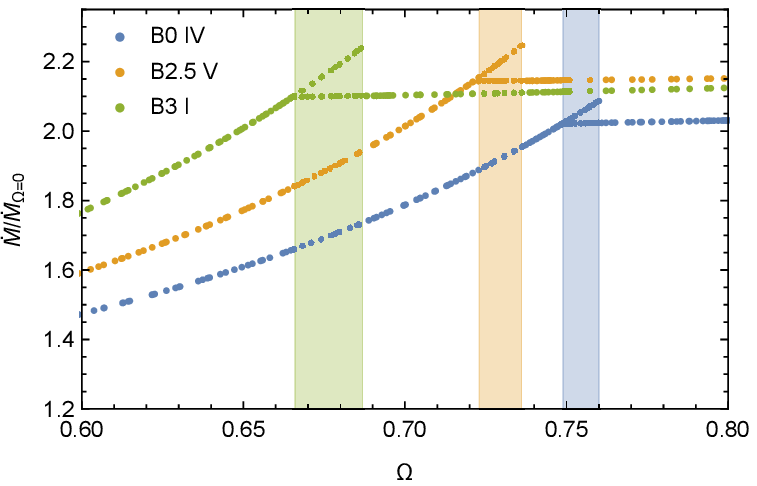}
	\vskip 0.3cm	
	\includegraphics[width=\columnwidth]{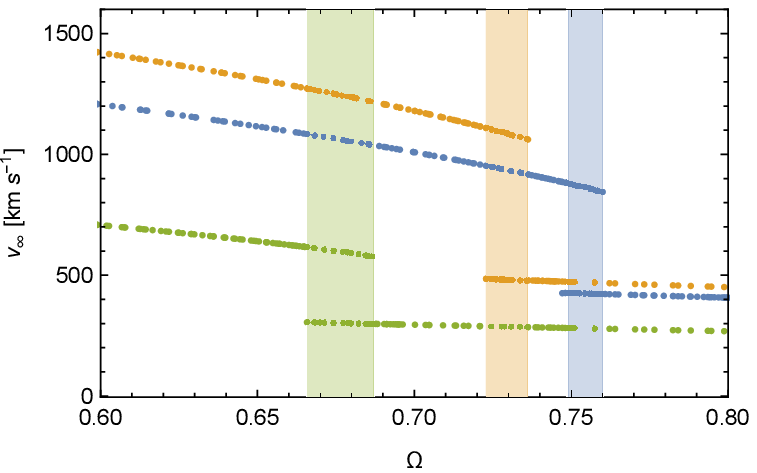}
    \caption{Similar to Figure \ref{stars-coex}, but with a zoom focuses on the regions where both type of solutions co-exist. This region is highlighted for each stellar model.}
    \label{stars-coex-zoom}
\end{figure}

The wind solutions of these three models are calculated with \textsc{Hydwind} for different values of $\Omega$. Their base densities are fixed to certain value $\rho_0$, which has the same value in each model for both types of solutions. These values are given in the last column of Table \ref{param}. 
The calculations begin with a  model with $\Omega=0$ and, then, the rotational speed is gradually increased until the last fast solution is attained. For $\Omega$-slow solutions a $\beta$-law with $\beta \approx 3.5$ and $v_{\infty}\approx 400$ km s$^{-1}$ is used as initial trial solution. To obtain the $\Omega$-slow solutions, we start with a model rotating at $\Omega=0.99$ and, then, $\Omega$ is gradually decreased until the last $\Omega$-slow solution is obtained.
The calculated mass-loss rates and terminal velocities as function of $\Omega$ are shown in Figure \ref{stars-coex}. We observe clearly the characteristic behaviour of the fast and $\Omega$-slow solutions. Furthermore, it is found that there exists an interval (or region) where fast and $\Omega$-slow solutions {\it{co-exist}}. 
The co-existence region of our models are zoomed in and highlighted in Figure \ref{stars-coex-zoom}. This co-existence region is different for each model and is roughly located in the interval $0.65\lesssim \Omega \lesssim 0.76$. In Appendix \ref{Apendice} we show our full topological analysis when $\Omega=0.73$. Table \ref{omin} summarizes the properties of the different solutions and lists the wind parameters  calculated at the minimum ($\Omega_1$) and maximum ($\Omega_2$) values  encompassed  by the co-existence region.  Overall, the terminal velocities of fast solutions are higher than the ones from the $\Omega$-slow solutions by factor of about two. On the other hand,  mass-loss rates computed for the fast and slow-$\Omega$ solutions at $\Omega_1$ are the same. While the maximum mass-loss rate of the fast solution, at the upper limit of the interval, $\dot{M}$($\Omega_2$), is not larger than $7\%$ with respect to the one calculated for the $\Omega$-slow solution.

\begin{table*}
\center
\caption{Properties of the fast and $\Omega$-slow solutions for our three stellar models. Wind parameters are calculated at values of $\Omega$ where the co-existence region begins and ends.}
\label{omin}
\tabcolsep 2.5pt
\begin{tabular}{llcccrccrc}
\hline
Model & $\Omega$ range & Solution type & Co-existence interval &$\dot{M}$($\Omega_1$) & $v_{\infty}$($\Omega_1$) & $r_{\mathrm{crit}}$($\Omega_1$)  &$\dot{M}$($\Omega_2$) & $v_{\infty}$($\Omega_2$) & $r_{\mathrm{crit}}$($\Omega_2$) \\
     &  &  &   $\Omega_1 - \Omega_2$ & [M$_{\sun}\,\mathrm{yr}^{-1}$] & [$\mathrm{km} \, \mathrm{s}^{-1}$]  & [R$_{*}$]    & [M$_{\sun}\,\mathrm{yr}^{-1}$] & [$\mathrm{km} \, \mathrm{s}^{-1}$] & [R$_{*}$] \\
\hline
B0\,IV  & $0-0.760$    & Fast     & $0.749-0.760$ &  2.44$\times10^{-7}$ & 880.4~~ & 1.056  & 2.52$\times10^{-7}$ & 844.5~~ & 1.058\\ 
          & $0.749-1$ & $\Omega$-slow & &2.44$\times10^{-7}$ & 426.7~~ & 16.32  &2.45$\times10^{-7}$ & 422.3~~ & 16.62 \\ 
\hline   
B2.5\,V & $0-0.736$ & Fast          & $0.723-0.736$ & 2.16$\times10^{-9}$ & 1109.3~~ & 1.027  & 2.26$\times10^{-9}$ & 1061.7~~ & 1.028\\  
       & $0.723-1$ & $\Omega$-slow && 2.15$\times10^{-9}$ & 485.1~~ & 22.36  & 2.15$\times10^{-9}$ & 478.7~~ & 22.36 \\
\hline   
B3\,I   & $0-0.687$ & Fast          & $0.666-0.687$ & 1.38$\times10^{-6}$ & 616.1~~ & 1.048          & 1.47$\times10^{-6}$ & 578.9~~ & 1.050\\  
       & $0.666-1$ & $\Omega$-slow & & 1.38$\times10^{-6}$ & 305.1~~ & 13.83  & 1.38$\times10^{-6}$ & 299.0~~ & 14.05 \\       
\hline
\end{tabular}
\end{table*}

\begin{figure}
	\includegraphics[width=\columnwidth]{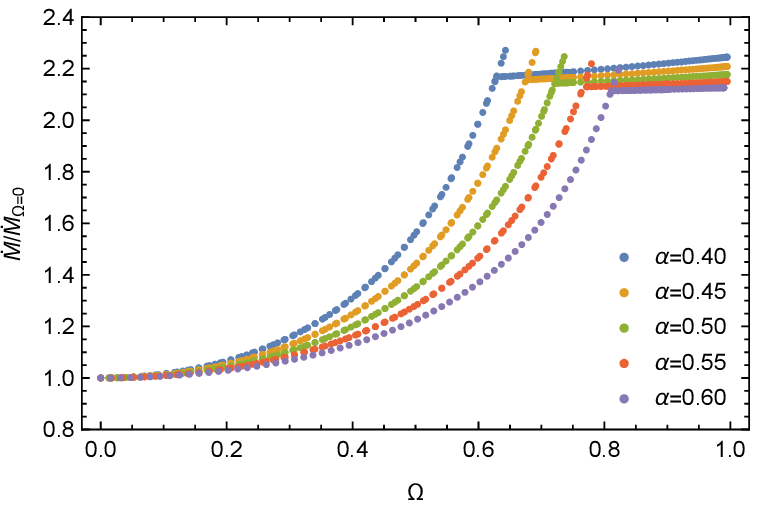}
	\vskip 0.3cm	
	\includegraphics[width=\columnwidth]{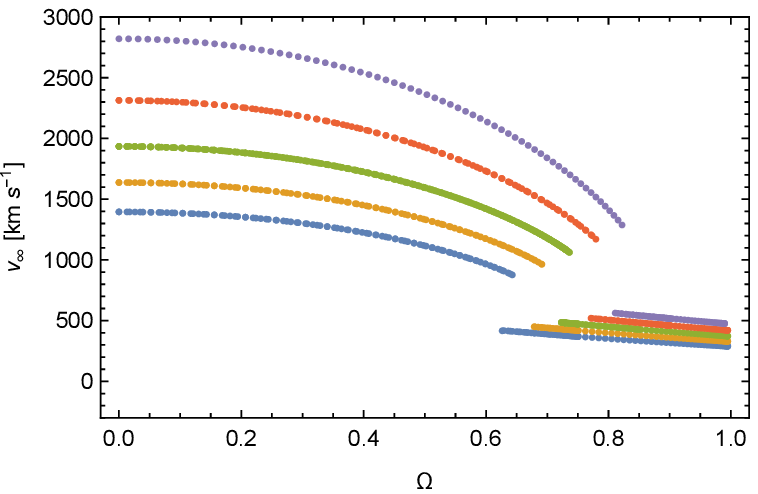}
    \caption{Mass-loss rates in units of a non-rotating case (top panel) and terminal velocities (bottom  panel) as a function of $\Omega$ for solutions obtained from the B2.5V model varying systematically the value of the line-force parameter $\alpha$. We observe a shift of the co-existence region towards larger rotational speeds as $\alpha$ increases.}
    \label{MOF07-alpha}
\end{figure}

\begin{figure}
	\includegraphics[width=\columnwidth]{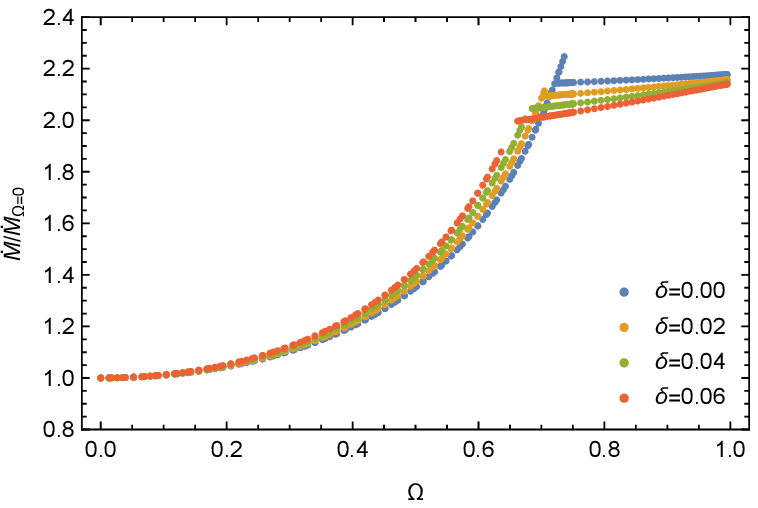}
	\vskip 0.3cm
	\includegraphics[width=\columnwidth]{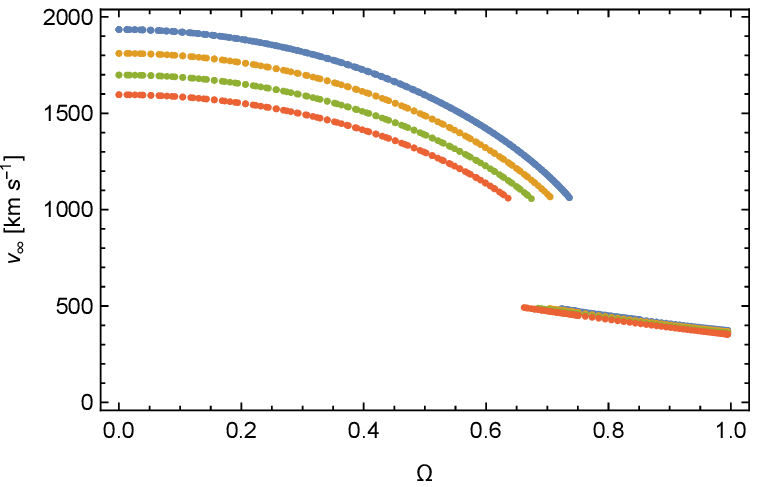}
    \caption{Similar to Figure \ref{MOF07-alpha}, but the models are calculated varying the value of the line-force parameter $\delta$. A gap between fast and $\Omega$-slow solutions is noted for the highest values of $\delta$.}
    \label{MOF07-delta}
\end{figure}

Finally, we study the influence of the line-force parameters on the co-existence region. In Figure \ref{MOF07-alpha} and \ref{MOF07-delta} we display mass-loss rates and terminal velocities as a function of $\Omega$, which are derived from the solutions obtained with \textsc{Hydwind} varying systematically the values of $\alpha$ and $\delta$ for the stellar model B2.5 V. As the value of $\alpha$ increases there is a `shift' of this region towards higher values of $\Omega$. This is similar to the results obtained by \citet{cure2004}, where for thin lines driving the wind (lower $\alpha$), lower rotational speeds are needed to obtain a $\Omega$-slow solution. Furthermore, the width of this region seems to be independent of the value of $\alpha$. Regarding $\delta$, the width of the co-existence region increases as $\delta$ increases below certain threshold, in this case $\delta \leq 0.04$.  For values larger than the threshold one, a gap is observed \citep{venero2016}, confirming the absence of monotonically increasing stationary solutions. \\

\section{1D time-dependent wind solutions for rotating stars}
\label{time_dep}

To study the stability of the steady state solutions presented in the previous section we solve here the time-dependent m-CAK model in the equatorial plane. For this purpose, we employ the hydrodynamic code \textsc{Zeus-3D} \citep{clarke1996}. We have adapted this Eulerian finite-difference code to our specific problem including the radiative driving terms along with all our assumptions (see Section \ref{approx}). 

Our spatial mesh has 500 zones going from $r_{1}=R_{*}$ to $r_{500}=100\,R_{*}$ distributed non-uniformly with higher concentration near the stellar surface to account for steeper flow gradients. Then, the zone spacing is defined by $\Delta r_{i+1}/\Delta r_{i}=1.02$, with $\Delta r_{i}=r_{i+1}-r_{i}$. Our test calculations have shown that this grid resolution is sufficient for our purpose and a grid with higher resolution will produce very similar results. 

At the inner boundary we set the base density to a factor of the sonic point density, $\rho(R_*)= f\,\rho(R_{s})$, in order to reach the stability of the outflow. We find that $f$ in the range of $5 <f <20$ gives us a stable solution. In addition, the inflow inner velocity boundary is allowed to be adjusted by a linear extrapolation from the closest zones in the interior of the computational domain. This boundary also is limited to a subsonic flow.  At the outer boundary we set  an outflowing wind as condition.

For numerical stability in the integration process, we follow \citet{madura2007} who truncate the radial velocity gradient to zero  whenever it is negative, i.e., $dv/dr \rightarrow \max \left( dv/dr,0 \right)$. This comes from a simple compromise between the likely extremes of $dv/dr$: (1) a lower limit where a negative velocity gradient implies a prior line resonance that shadows radial photons from the star, and (2) an upper limit where this shadowing can be weaken by forward scattering \citep[see more details in][]{madura2007}. 

In order to begin the integration of the time-dependent hydrodynamic equations for rotating winds with \textsc{Zeus-3D}, we start with the stationary solutions obtained by \textsc{Hydwind} as initial or trial solution. Moreover, we also performed calculations using various $\beta$-velocity field trial solutions, 
as we did for the stationary cases in the previous section. 

\begin{figure}
	\includegraphics[width=\columnwidth]{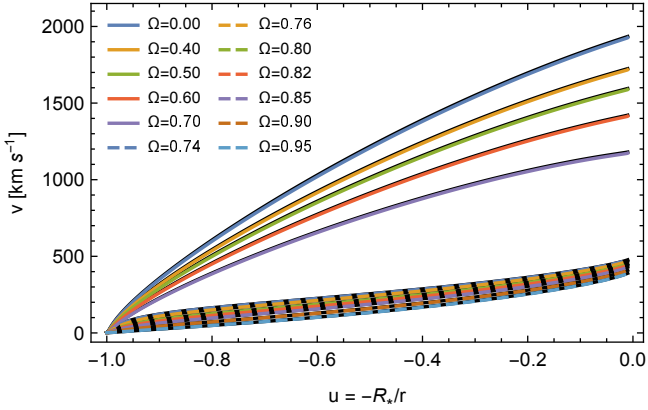}
    \caption{Velocity profiles obtained with the code \textsc{Zeus-3D} for different values of $\Omega$ for  B2.5\,V model. As initial condition a stationary solution for each value of $\Omega$ is used. Solid and dashed lines represent the fast and slow solutions, respectively. The time-dependent solutions are compared with the stationary solutions (black solid lines overlaid by the time-dependent solutions).}
    \label{mof-zeus-profiles}
\end{figure}

Figure \ref{mof-zeus-profiles} shows the velocity profiles for different values of $\Omega$  obtained with \textsc{Zeus-3D}  for the model B2.5\,V (see Table \ref{param}), we also overplotted the  stationary solutions shown in Figure \ref{mof-hyd-profiles}. The agreement between both results, using completely different numerical algorithms, is excellent. 
The relative error with respect to the stationary solution for  terminal velocities is about $1\%$, while the relative error in the mass-loss rate is less than $12\%$. The disagreement in the mass-loss rate is mainly due to a difference between the density profiles. This variance arises from the use of two codes (time-dependent and stationary) which employ different numerical  approaches for different equations of motion.

	It is important to stress here that our results show that there are no kinks in the resulting velocity profiles calculated with time-dependent solutions (see Figure \ref{mof-zeus-profiles}), in disagreement with the results found by \citet{madura2007}. 

\begin{figure}
	\includegraphics[width=\columnwidth]{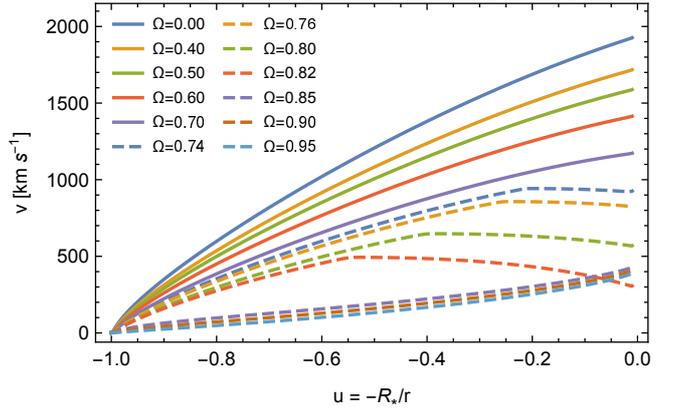}
	\vskip 0.3cm
	\includegraphics[width=\columnwidth]{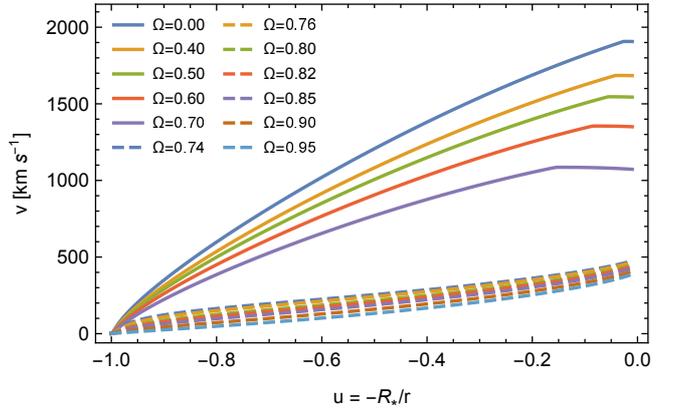}
	\caption{Velocity profiles obtained with the code \textsc{Zeus-3D} at different values of $\Omega$ for the B2.5\,V model. {\it Top panel:} As initial condition  we used a stationary solution with $\Omega=0$ from \textsc{Hydwind}. We obtained fast and $\Omega$-slow solutions but, now, models with $\Omega=0.74$, 0.76, 0.80 and 0.82 present a kink, similar to the velocity profiles obtained by \citet{madura2007}. {\it Bottom panel:}  As initial condition we adopted a stationary solution with $\Omega=0.95$ from \textsc{Hydwind}. The $\Omega$-slow solutions are recovered but all the fast solutions present a kink.}
    \label{mof-zeus-kink1}
\end{figure}

In order to reproduce \citet{madura2007} results, we calculated  time-dependent solutions  employing first an initial {\it{non-rotating}} trial solution as these authors  had done. However, in our case we started from the solutions obtained by \textsc{Hydwind}. These results are shown in the top panel of Figure \ref{mof-zeus-kink1} where we can now notice the presence of kinks in the velocity profiles for $\Omega = 0.74$, 0.76, 0.80, and 0.82, in agreement with the calculations performed by \citet{madura2007}. These kinks occur far from the stellar surface due to overloading of the wind and become more prominent as the value of $\Omega$ increases. It is worth mentioning that these kink-solutions should be the $\Omega$-slow solutions shown  in Figure \ref{mof-zeus-profiles}, in dashed lines. 

Second, we repeat these calculations, but now using as initial trial solution a stationary solution with $\Omega=0.95$ from \textsc{Hydwind}. The bottom panel of Figure \ref{mof-zeus-kink1} shows the different velocity profiles. Here we can notice that the velocity profiles from the $\Omega$-slow solutions do not present any kink, whereas the velocity profiles of all fast solutions do. These  kinks follow the same behaviour as in our previous calculations, i.e., the lower is the rotational speed the farther the kink occurs from the stellar surface.


We would like to remark that stationary solutions with a kink (attained with \textsc{Zeus-3D}) use the truncation condition $dv/dr \rightarrow \max \left( dv/dr,0 \right)$ from the kink location up to the outer boundary. Instead, models without kink have a positive velocity gradient along the whole  velocity profile so the truncation condition is not used.

We conclude that kinks arise when 
we use  an  initial (trial) solution that belongs to a different solution branch in the topology of the m-CAK model  \citep[see topological analysis by ][]{cure2007}.


\subsection{Co-existence of fast and slow regimes}

We wish now to examine whether the time-dependent \textsc{Zeus-3D} code confirms  the co-existence of  solutions  found in Section \ref{coexistence}. For this purpose, we calculated time-dependent solutions for the 3 models shown in  Table \ref{param}. We found almost the same $\Omega$-interval of co-existence, with a difference of the order of $\sim 10^{-3}$ for $\Omega_1$ and $\Omega_2$.

\begin{figure}
	\centering
	\includegraphics[width=\columnwidth]{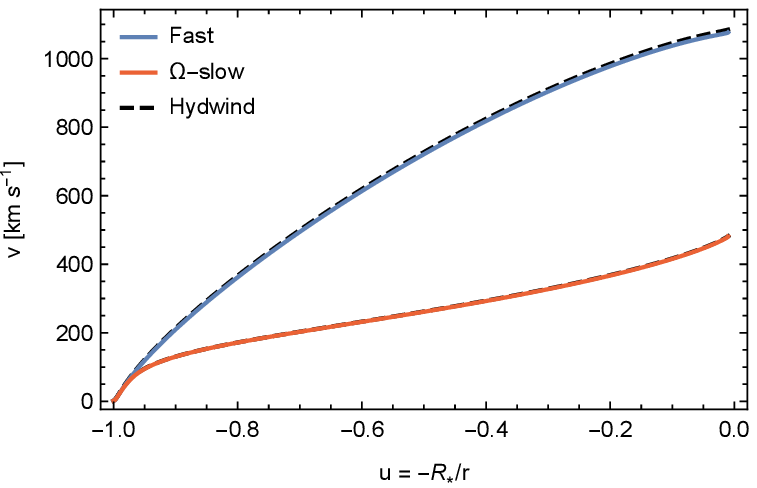}
	\vskip 0.3cm	
	\includegraphics[width=\columnwidth]{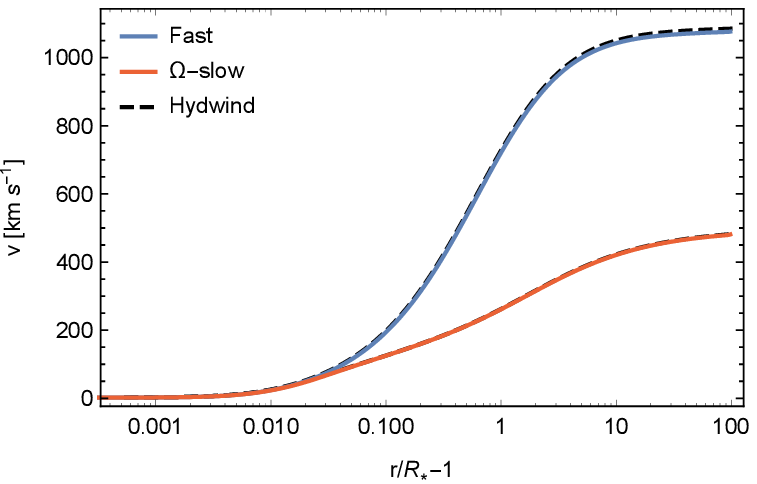}
	\vskip 0.3cm	
	\includegraphics[width=\columnwidth]{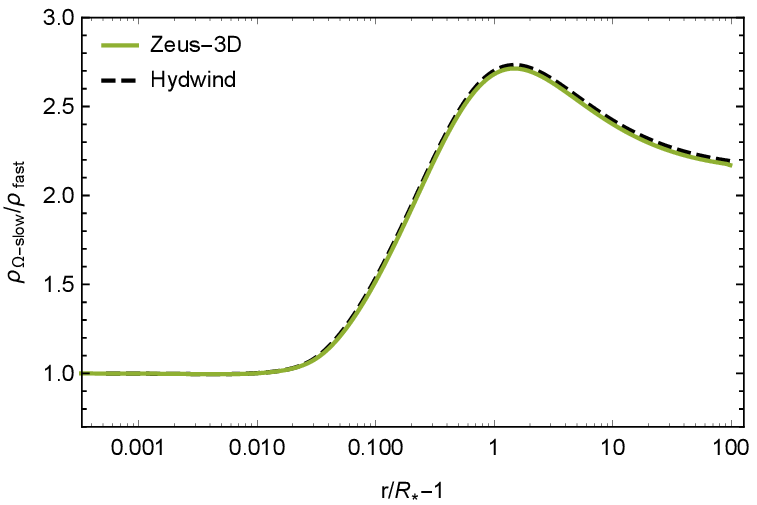}
        \caption{ 
        	{\it Top panel:} Velocity profiles as function of the inverse radial coordinate $u$.  Blue and red solid lines represent the time-dependent fast and $\Omega$-slow solutions, respectively, for $\Omega=0.75$. Black dashed lines are the corresponding stationary solutions. {\it Middel panel}: Same as the top panel but now as function of $r/R_{*}-1$.
        	{\it Bottom panel:} Density ratio between time-dependent fast and  $\Omega$-slow solutions as function of $r/R_{*}-1$. The stationary solutions are also shown in black dashed lines and are overlaid with the time-dependent ones.}
    \label{mof07-coex}
\end{figure}

Figure \ref{mof07-coex} (top and middle panels) shows the results for the  B2.5\,V model for $\Omega=0.75$. Both time-dependent velocity profiles match very closely their corresponding stationary solutions (black dashed lines). Bottom panel illustrates the density ratio as function of $r/R_{*}-1$, for the time dependent (green solid line) and stationary (black dashed line) solutions.


\section{Switch between fast and slow velocity regimes}

With the purpose of studying the stability of the solutions in the co-existence region and a possible switch of wind regime, we performed
density perturbations at the base of the wind using \textsc{Zeus-3D}.

\par
We conduct simulations using three different types of density perturbations as function of time: (1) a random perturbation where the perturbed base density,  $\rho_{\mathrm{pert}}$, takes random values around the steady-state base density, $\rho_{0}$. The amplitude of the perturbation is expressed in terms of $\rho_{\mathrm{pert}}/\rho_{0}$; (2) a square  perturbation where the value of $\rho_{0}$ switches from a fixed upper (or lower) value $\rho_{\mathrm{pert}}$; and (3) a Gaussian perturbation where $\rho_{\mathrm{pert}}$ increases (decreases) with respect to $\rho_{0}$, following a Gaussian profile. 

\begin{figure}
	\includegraphics[width=\columnwidth]{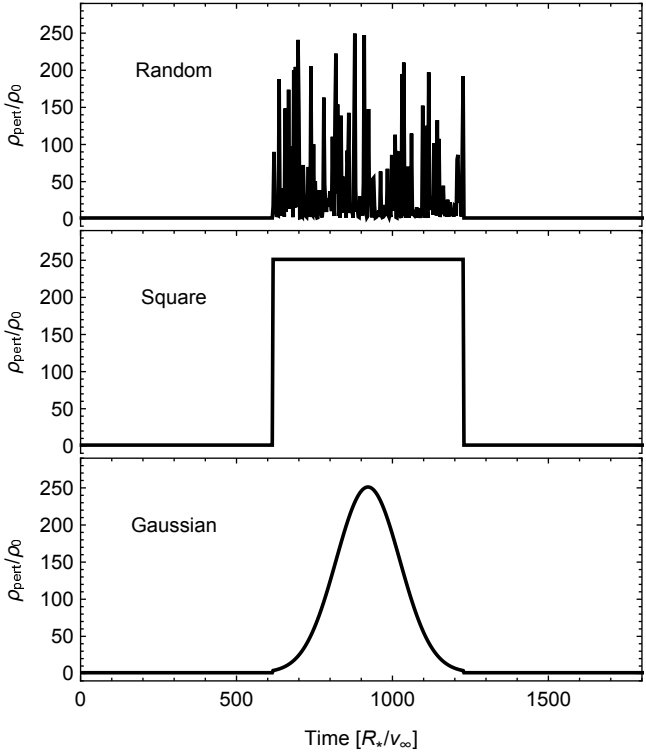}
    \caption{Typical shapes of perturbations with enhanced base density. The enhanced base density is in units of base density $\rho_{0}$ and the time is in units of the flow time $t=R_{*}/v_{\infty}$. The perturbation time is $1000$ ks.
    \label{pert-type}}
\end{figure}

Figure \ref{pert-type} illustrates the behaviour of the three different types of perturbations of the base density with the same enhanced amplitude and time. Time is given in units of the flow time $t=R_{*}/v_{\infty}$. 
Random perturbations represent sudden change; square perturbations obey  a strong and sudden change and Gaussian perturbations correspond to a smooth and continuous change in density at every instant.

Based on these perturbations we searched for the minimum amplitude ($\rho_{\mathrm{pert}}/\rho_{0}$) and minimum time that could lead to a switch between a fast to a slow regime, and vice versa.

\par
From the numerical simulations we performed, we conclude that we need to increase the base density, $\rho_0$, to switch from a fast solution to a $\Omega$-slow one and reduce $\rho_0$ for the other case. Tables \ref{pert-t1} and \ref{pert-t2} show the results of our simulations for each switching case. These tables also show the minimum amplitudes and time required to switch regime. All these perturbations last a period of time which is typically about 500 flow times for transitions from fast to $\Omega$-slow solutions and about 100 flow times for transitions from $\Omega$-slow to fast solutions. These period of times are similar to the time relaxation from the initial trial solution to a steady state. 

In the case of random perturbations, the model B2.5 V requires a much higher density amplitude perturbation than the other models to switch from fast to $\Omega$-slow regime. Nevertheless, for models B2.5 V and B3 I, square and Gaussian density perturbations do not switch at all, and the B0 IV model requires an unrealistic amplitude perturbation. When the regimes do not switch at all, a kink structure results in most of the studied cases. 

When the $\Omega$-slow solution is perturbed, the switch to the fast solutions is attained by lowering the density by a factor  $\lesssim 1/100$ for square and Gaussian perturbations for all models. For random perturbations, the B0 IV model never switches to the fast solution and the other two models require  very high amplitudes, e.g., B2.5 V model requires an amplitude factor  $\approx 1/1000$ to switch. Based on the physically unlikely high values required to switch from one solution to another, both type of solutions seem very stable. However, is more likely a switch from a $\Omega$-slow to fast solution.

\begin{table}
\caption{Minimum conditions required from a density enhanced perturbation to switch from fast to $\Omega$-slow solution within the co-existence region. The models with no data correspond to simulations where under none condition a switch of solutions is obtained.}
\label{pert-t1}
\begin{tabular}{lccl}
\hline
Star & Time & Amplitude  & Type\\
     & (s) &  $\left(  \rho_{\mathrm{pert}}/\rho_{0} \right)$ &  \\
\hline
B2.5 V          & $6 \times 10^{6}$  & $1 \times 10^{4.8}$ & Random \\
                    & --  & -- &        Square  \\
                    & --  & -- &        Gaussian  \\                          
\hline
B0 IV      & $1 \times 10^{7}$     & $1 \times 10^{2.4}$ & Random \\
              & $4 \times 10^{6}$  & $ 1 \times 10^{6}$ & Square  \\
              & $4 \times 10^{6}$  & $ 1 \times 10^{6}$ & Gaussian \\  
\hline
B3 I      & $5 \times 10^{7}$  & $1 \times 10^{2.7}$ & Random \\
&  --  &  --  & Square  \\
&  --  &  --  & Gaussian \\  
\hline

\end{tabular}
\end{table}

\begin{table}
\caption{Similar to Table \ref{pert-t1}, but here the switch is from a $\Omega$-slow to a fast solution with a decreased density perturbation.}
\label{pert-t2}
\begin{tabular}{lccl}
\hline
Star & Time & Amplitude & Type\\
     & (s)    & $\left(  \rho_{\mathrm{pert}}/\rho_{0} \right)$ &  \\
\hline
B2.5 V    & $5 \times 10^{5}$ & $1 \times 10^{-2.9}$ & Random  \\
              & $1 \times 10^{4}$ & $1 \times 10^{-2.0}$ & Square \\
              & $5 \times 10^{4}$ & $1 \times 10^{-2.0}$ & Gaussian\\                          
\hline
B0 IV      & --  & -- & Random \\
              & $2 \times 10^{5}$ & $1 \times 10^{-2.0}$ & Square  \\
              & $9 \times 10^{5}$ & $1 \times 10^{-2.0}$ & Gaussian \\  
\hline
B3 I      & $5 \times 10^{6}$ & $1 \times 10^{-5.5}$ & Random \\
& $3 \times 10^{6}$ & $1 \times 10^{-2.0}$ & Square  \\
& $3 \times 10^{6}$ & $1 \times 10^{-2.0}$ & Gaussian \\  
\hline
\end{tabular}
\end{table}

The perturbation time required to switch regimes are quite different. To obtain a $\Omega$-slow solution, starting from a fast solution, it requires a perturbation time of the order of $4 \times 10^{6}$ s -- $5 \times 10^{7}$ s or 500 -- 600 flow times. For the other switching case the time is of the order of $10^{4}$ s -- $5 \times 10^{6}$ s or a few -- 100 flow times. This difference is explained by the characteristic flow times of both regimes to achieve a state state from an initial trial solution.

\begin{figure}
	\includegraphics[width=\columnwidth]{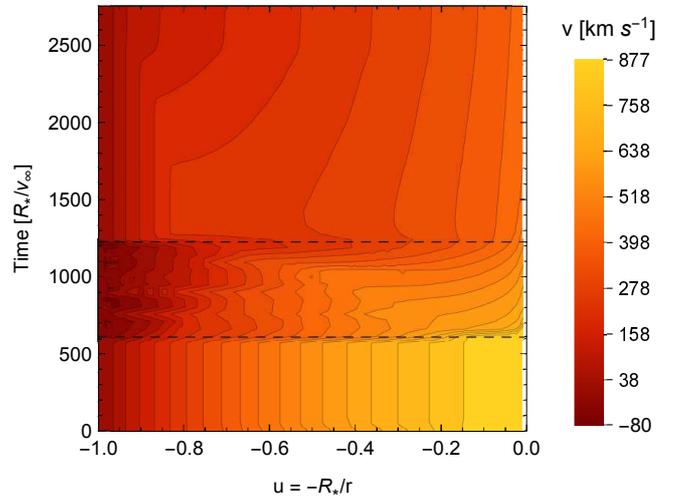}
    \caption{Velocity contours as function of time for the B0 IV model. Time is in units of the flow time $t=R_{*}/v_{\infty}$. The simulation begins from a fast solution  with $\Omega=0.75$. Then a random enhanced density perturbation with amplitude $10^{2.4}$ is applied and a switch to a $\Omega$-slow solution is attained. Dashed lines indicate the initial and final times of the perturbation process.}
    \label{gaCas-CP-F2S}
\end{figure}

\begin{figure}
	\includegraphics[width=\columnwidth]{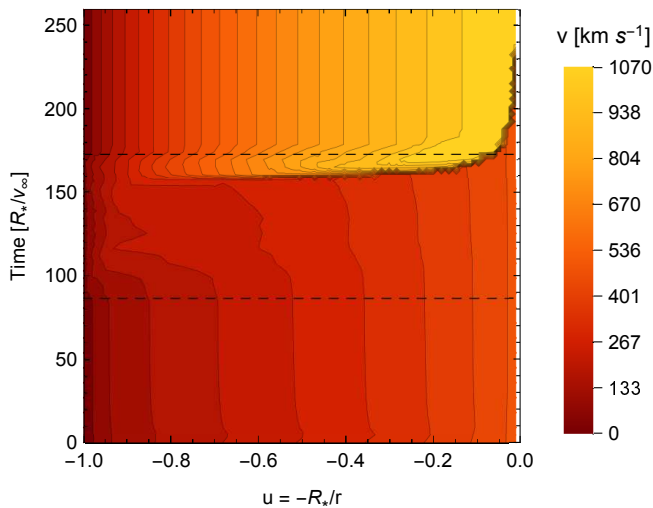}
    \caption{Same as Figure \ref{gaCas-CP-F2S}. This simulation corresponds to a B2.5 V model, starting with a $\Omega$-slow solution rotating at $\Omega=0.73$. A random decreased density  perturbation with amplitude $10^{-2.9}$ is applied and a switch to a fast solution is attained. 
}
    \label{MOF07-CP-S2F}
\end{figure}

Density and velocity contour-plots, starting from a fast regime is shown in Figure \ref{gaCas-CP-F2S} for the B0 IV model, while starting from $\Omega$-slow regime is shown in \ref{MOF07-CP-S2F} for the B2.5 V model. In each case we show the results when random perturbation were applied at the base density under the conditions given in Tables \ref{pert-t1} and \ref{pert-t2}. In both cases the solutions switches its regime.

\section{Discussion }

We studied in previous sections different density perturbations that could yield a switch of wind regime in the co-existence region. In this section we want to discuss the implications of this regime switching in the frame of variability of massive stars and their winds.

There exists certain consensus \citep{puls2008} that line-profile variability is induced due to an interplay between disturbances
in the photosphere, such as non-radial pulsations, stellar spots, etc. Thus, any mechanism (or a combination of them) that induces a significant change in the photosphere density when the stellar rotational speed is in the co-existence range ($\Omega_1\leq \Omega\leq\Omega_2$), might have a dramatic impact in the equatorial plane triggering a switch of wind regime. This could lead to the formation of a circumstellar disk in the equatorial plane of these stars, where the mass injection mechanism is given by radiation pressure in a $\Omega$-slow wind regime. These disks might appear at rotational speeds of the order of $\Omega \sim 0.65\,-\,0.75$, values that are much lower than the critical rotational speed. Moreover, these rotational velocities are in better agreement with the observed mean value from the distribution of rotational speeds of Be stars \citep[][and references therein]{zorec2016}.
If subsequent perturbations do not have the sufficient amplitudes, this scenario is stable in the frame of this 1D model. Otherwise, if the amplitudes of perturbations are high enough, there could be another switch to the fast regime, leading to a {\it{dissipation}} of the disk. These new configuration can also be stable for a long period of time. 

These regime switching might also explain the change of phases between B and Be stars and also episodic mass loss events in Be stars \citep{balona2010}. 

Linking this regime-switch with stellar evolution, as the internal structure of a star evolves, the critical rotational speed decreases \citep[$\Omega$ increases,][]{langer1998} and $\Omega$ might reach the co-existence region, creating-dissipating a disk. In a subsequent evolutionary scenario
characterized by $\Omega > \Omega_2$, only the $\Omega$-slow regime wind is present {at the equatorial plane} \citep{cure2004} leading to a different evolutionary phase with a higher mass-loss rate. Whenever a $\Omega$-slow regime is established, \citet{araya2017} have demonstrated that the wind stationary density structure is able to explain the observed H$\alpha$ emission lines in Be stars. 

In contrast to some radiation-driven viscous disk models \citep{kee2016,krticka2011} that need near critical rotational speed to create-dissipate a disk, our model requires under-critical rotation.

A current weakness of this m-CAK model is that it does not consider yet the role of viscosity and its influence on angular momentum transport, mechanism that might explain a Keplerian disk. Furthermore, it is important to emphasize, that the models presented here are purely 1D and solutions in multi-dimensions might be quite different from what we present here. Thus, a more extensive study considering a 2D/3D model with non-radial forces (and the effects of stellar distortion and gravity darkening) is required to confirm the co-existence region.

\section{Conclusions}
The topology of non-linear m-CAK differential equation predicts two wind solutions (from different solution branches) as function of $\Omega$ for a given set of line-force parameters. The fast solution ceases to exist when $\Omega$ is about certain threshold value ($0.65 \lesssim \Omega \lesssim 0.76	$) that depends on the stellar and line-force parameters, while the $\Omega$-slow solution begins to exist from this interval up to $\Omega \lesssim 1$. 

We investigated the region in the $\Omega$-space where these solutions are simultaneously present by solving 1D stationary and 1D time-dependent hydrodynamic equations at the equatorial plane. From the steady-state study we found a co-existence region inside a small interval of $\Omega$, where both solutions are {\it{simultaneously}} present, both satisfying the same boundary conditions at the stellar surface ($\rho(R_*)=\rho_0$).

We demonstrated that time-dependent solutions are very sensitive to the initial solution topology. If the topology of the initial velocity profile is far from the asymptotic steady state (belonging to a different solution branch), the calculation leads to 
nonmonotonic 'kink' solutions. Therefore, to obtain either fast or $\Omega$-slow solutions with a globally monotonic acceleration, we need to use an initial solution representative of a fast or slow regime, respectively. Our calculations confirmed the stability of these known stationary solutions and when using the proper initial conditions these solutions present no-kinks.

Concerning the location and width of a co-existence region, they depend on stellar and line-force parameters. Higher values of $\alpha$ shift the location of this region towards higher values of $\Omega$, while the width of the region is almost unchanged. 
For some particular values of $\delta$, we find a gap in the steady state solutions, confirming the results of \citet{venero2016}. 

We studied in the co-existence region, which minimum conditions can induce a switch of regime by means of density perturbations at base of the wind. We use namely: random, square and Gaussian perturbations with different amplitudes and perturbation times. Our results indicate that the switching processes are not identical, while a switch from fast to $\Omega$-slow regimes requires a perturbation amplitude of some hundreds, the reverse case requires a hundredth of the wind base density.  

Our main interpretation of this work is that the switching process can trigger the formation and/or dissipation of an outflowing equatorial disk
without needing a star rotating at almost critical rotational speed.

\section*{Acknowledgements}
The authors would like to thank the referee, Achim Feldmeier, for his thoughtful comments and suggestions. We also thank Diego Rial for his helpful comments concerning the topological analysis. I.A. acknowledges support from Fondo Institucional de Becas FIB-UV. M.C. and I.A. acknowledges support from Centro de Astrof\'isica de Valpara\'iso. 
A.uD acknowledges support by NASA through Chandra Award numbers GO5-16005X, AR6-17002C, G06-17007B and TM7-18001X issued by the Chandra X-ray Observatory 
Center which is operated by the Smithsonian Astrophysical Observatory for and behalf of NASA under contract NAS8- 03060. A. ud-Doula was supported by the FRS-FNRS at the University of Li\`ege (Belgium) for a research stay. L.C. acknowledges financial support from CONICET (PIP 0177), La Agencia Nacional de Promoci\'on Cient\'ifica y Tecnol\'ogica (PICT 2016-1971) and the Programa de Incentivos (G11/137) of the Universidad Nacional de La Plata (UNLP), Argentina. L.C. and M.C. thank support from the project CONICYT + PAI/Atracci\'on de capital humano avanzado del extranjero (folio PAI80160057).



\bibliographystyle{mnras}
\bibliography{cites} 


\appendix
\section{Topological Analysis}
\label{Apendice}

In this section we present a topological analysis of a rotational radiation driven wind, this analysis is based in the work
of \citep{cure2007}, where we presented in detail all the topological theory of radiation driven winds. 
Here we present a brief summary and the reader might refer to \cite{cure2007} for more details.
\subsection{Topology of singular points}
We define the logarithmic variables $\eta$ and $\zeta$ in terms of $u$, $w$ and $w'$ as follows:
\begin{equation}
\eta = \ln(2 w' / w)
\end{equation}
and 
\begin{equation}
\zeta = \ln(w w')
\end{equation}
Then, from the equation of motion (Eq. \ref{motion-eq}) plus the regularity and singularity conditions at the singular point, we were able to
solve for $\zeta=\zeta(u,\eta,C')$ and then define two new functions,  $H(u,\eta,C')$ and  $R(u,\eta)$  \citep[see Section 3.4 from][]{cure2007}. 
The location of a singular point is when the following conditions are simultaneously satisfied:
\begin{equation}
H(u,\eta,C')\,=\, R(u,\eta)\,=\,0
\end{equation}

\begin{figure}
	\includegraphics[width=\columnwidth]{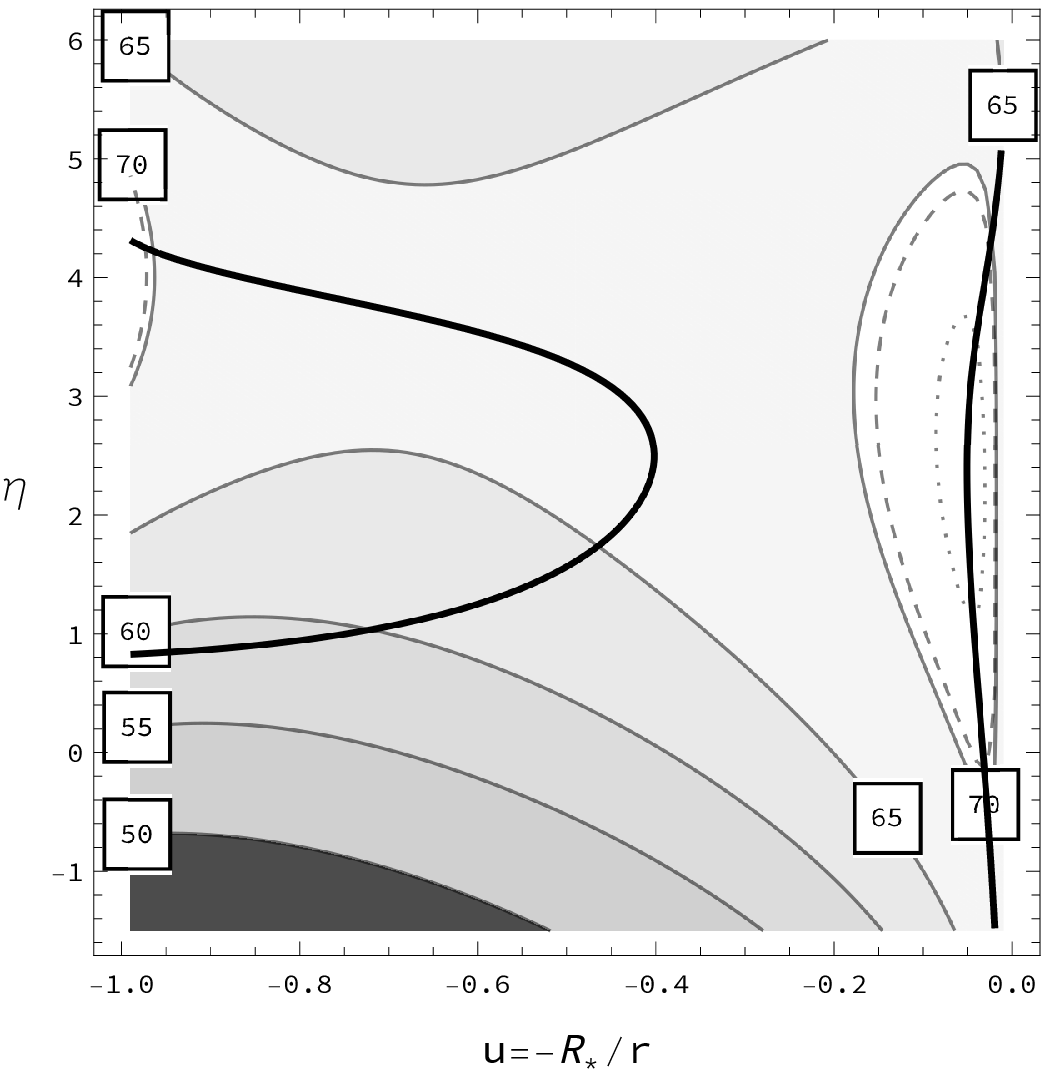}
    \caption{ $H(u,\eta,C')$ and  $R(u,\eta)$ functions in terms of $u$, $\eta$ and $C'$, see text for details}
    \label{figHR}
\end{figure}

Figure \ref{figHR} shows $H(u,\eta,C')$ and  $R(u,\eta)$ functions in terms of $u$, $\eta$ and $C'$.
Dark-gray continuous lines show the contour plots $H(u,\eta,C')=0$ for different values of the eigenvalue $C'$ (Square labeled). 
In addition the dashed contour lines are the $H(u,\eta,C')=0$ function for $C'=70.37$, 
which is the eigenvalues of the {\it{fast solution}}. 
The contour line corresponding to eigenvalue $C'=70.37$ has two components: one on the left of left component
of contour line of $C'=70$ and the other inside the closed contour 
of $C'=70$ at the right border of Figure \ref{figHR}.
The dotted closed line corresponds to level line of $H(u,\eta,C')=0$ 
for $C'=71.32$, which is the eigenvalue of the {\it{$\Omega$-slow solution}}.
The function $R(u,\eta)=0$ is also plotted in this figure in solid thick black lines. 
Functions $H(u,\eta,C')$ and  $R(u,\eta)$ intersects in five different locations for both eigenvalues, 
three for the fast solution eigenvalue and two for the $\Omega$-slow solution eigenvalue. 
All singular point are labeled from A to E clockwise starting from the Fast solution eigenvalues as shown in Table \ref{topologies}. 
This table also show the location of the each  singular point in the $u,\eta$ plane and the value of the determinant of matrix $B$, 
that give us the information concerning the topology type of a singular point  \citep[see definition of matrix $B$ in section 5.2 from][]{cure2007}. In all these cases,  the determinant values were negative, thus all singular points are X-type.

\begin{table}
\caption{Locations and topology of the singular points}
\label{topologies}
\begin{tabular}{lccccc}
\hline
Label  & $C'$  & $u$& $\eta$ & det($B$) & Topology   \\
\hline
A (fast)& 70.37 & -0.97327 & 4.2516 & $<0$ & X-type  \\
B & 70.37 & -0.02579 & 4.2221 & $<0$ & X-type  \\
C & 71.32 & -0.04045 & 3.5097 & $<0$ & X-type  \\
D (slow)& 71.32 & -0.04396 & 1.1126 & $<0$ & X-type  \\
E & 70.37 & -0.03142 &-0.1433 & $<0$ & X-type  \\
\hline
\end{tabular}
\end{table}

\subsection{Integration from  singular points}
\begin{figure}
	\includegraphics[width=\columnwidth]{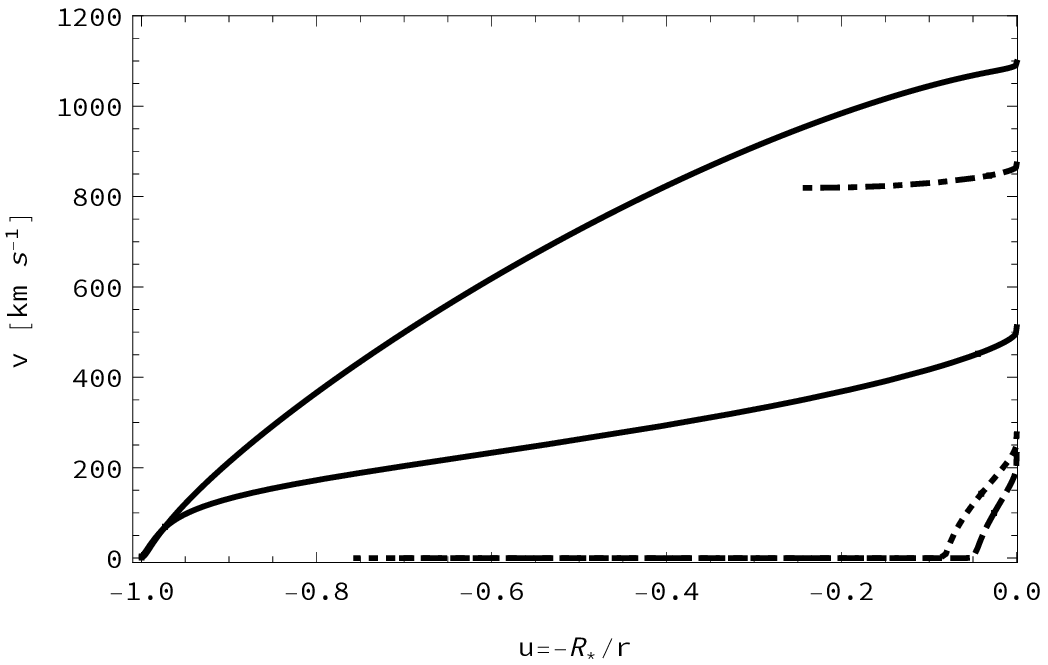}
    \caption{Velocity profile as function of $u$. We start from any singular point and integrate up and downstream. Only fast and $\Omega$-slow solutions (in solid lines) reach the stellar surface. All other solutions have no physical meaning because never reach the stellar surface.}
    \label{sols}
\end{figure}

Consequently, we start to integrate from each singular point outwards and inwards to obtain the velocity profile as function of $u$.
Figure \ref{sols} show the velocity profiles of each of the integrations.

Fast (A) and $\Omega$-slow (D) solutions are plotted in black solid lines, the comparison with Figure \ref{mof07-coex} is remarkable. The integration from singular points B, C and E are plotted in dashed, dotted and dot-dashed lines respectively. No one of these velocity profiles reach the stellar surface, so we conclude that these are non-physical solutions of our rotating radiation driven wind.

Finally we want to remark that both physical solutions (fast \& $\Omega$-slow) founded by \textsc{Hydwind} \& \textsc{Zeus-3D} have the same initial condition at stellar surface, rotational speed, stellar and line force parameter and the stationary solutions have different location of the singular point and {\it{different eigenvalues}}.


\bsp	
\label{lastpage}
\end{document}